\documentclass[runningheads]{llncs}

\usepackage{amssymb}

\usepackage{xcolor}
\usepackage{graphicx}
\usepackage{textcomp}
\usepackage{amsmath}
\usepackage{array}
\usepackage{multirow}
\usepackage{url}
\usepackage[ruled,vlined]{algorithm2e}
\usepackage{cite}
\usepackage{hyperref}
\hypersetup{
    colorlinks=true,
    linkcolor=blue,
    urlcolor=blue,
    citecolor=blue,
    pdfborder={0 0 0} 
}
\begin{document}
\title{A Novel Low-Rank Tensor Method for Undersampling Artifact Removal in Respiratory Motion-Resolved Multi-Echo 3D Cones MRI}
\titlerunning{Low-Rank Tensor Method for Artifact Removal in Motion-Resolved MRI}

\author{Seongho Jeong$^{1,2}$, MungSoo Kang$^{1}$, Gerald Behr$^{1}$, Heechul Jeong$^{2}$, and Youngwook Kee$^{1}$}
\authorrunning{S. Jeong et al.}
%
\institute{$^1$Memorial Sloan Kettering Cancer Center, New York, NY, USA \\
$^2$Kyungpook National University, South Korea}
\institute{$^1$Memorial Sloan Kettering Cancer Center, New York, NY, USA\\
$^2$Kyungpook National University, South Korea \\ \email{\{jeongs1, kangm2, behrg, keey\}@mskcc.org; \{setdotpy, heechul\}@knu.ac.kr}}
\maketitle              
\vspace{-0.2cm}
\begin{abstract}
We propose a novel low-rank tensor method for respiratory motion-resolved multi-echo image reconstruction. The key idea is to construct a 3-way image tensor (space $\times$ echo $\times$ motion state) from the conventional gridding reconstruction of highly undersampled multi-echo k-space raw data, and exploit low-rank tensor structure to separate it from undersampling artifacts. Healthy volunteers and patients with iron overload were recruited and imaged on a 3T clinical MRI system for this study. Results show that our proposed method Successfully reduced severe undersampling artifacts in respiratory motion-state resolved complex source images, as well as subsequent R2* and quantitative susceptibility mapping (QSM). Compared to conventional respiratory motion-resolved compressed sensing (CS) image reconstruction, the proposed method had a reconstruction time at least three times faster, accounting for signal evolution along the echo dimension in the multi-echo data.
\keywords{Low-Rank Tensor Decomposition \and 3D Non-Cartesian MRI \and Motion-Robust Quantitative Imaging.}
\end{abstract}
%
\section{Introduction}

Respiratory motion is a major challenge in abdominal MRI, often requiring multiple breath-holds from the subject \cite{feinberg1995multiple}. This challenge is exacerbated in pediatric patients and some adults with medical conditions \cite{kee2021free}. Recent advances in 3D multi-echo gradient-echo non-Cartesian MRI with respiratory motion-resolved compressed sensing (CS) reconstruction have enabled motion-robust, fully ungated, and free-breathing quantitative body imaging, including proton density fat fraction (PDFF), R2*, and quantitative susceptibility mapping (QSM) \cite{armstrong2018free, schneider2020free, kee2021free, kang2022ismrm}. The accuracy of these quantitative tissue parameter maps has improved over conventional motion-averaged reconstruction with respect to the clinical standard Cartesian MRI with successful breath-holds.
Yet, there are two remaining challenges with the current state-of-the-art CS-based motion-resolved reconstruction in 3D multi-echo MRI from a computational imaging perspective. First, the MRI signal evolution along the echo dimension is not accounted for in the reconstruction, which may result in unwanted reconstruction artifacts and lead to unreliable or inaccurate PDFF, R2*, and QSM. 
Second, a significant memory footprint and the use of forward/backward 3D nonuniform Fourier transform (NUFFT) result in a computational bottleneck. This bottleneck is more pronounced in full 3D non-Cartesian k-space trajectories, such as 3D cones, as opposed to stack-of-radial/spiral trajectories, where slices in the stacking dimension can be reconstructed in parallel.
In this paper, we propose a low-rank tensor method based on the CANDECOMP/PARAFAC Decomposition (CPD) \cite{kolda2009tensor, sidiropoulos2017tensor} with total variation (TV) \cite{rudin1992nonlinear} regularization to address the computational challenges mentioned above. The key idea is that the proposed tensor formulation of the 5D imaging data (3D space +  1D echo evolution + 1D motion state) and its low-rank tensor approximation effectively remove severe undersampling artifacts from the simple gridding reconstruction of each motion state. This does not require successive forward/backward 3D NUFFTs. As a result, the proposed method offers faster reconstruction, leaves a smaller memory footprint, and accounts for signal evolution along the echo dimension.

\section{Theory}

\begin{figure}[t!]
\centering
\includegraphics[width=\textwidth]{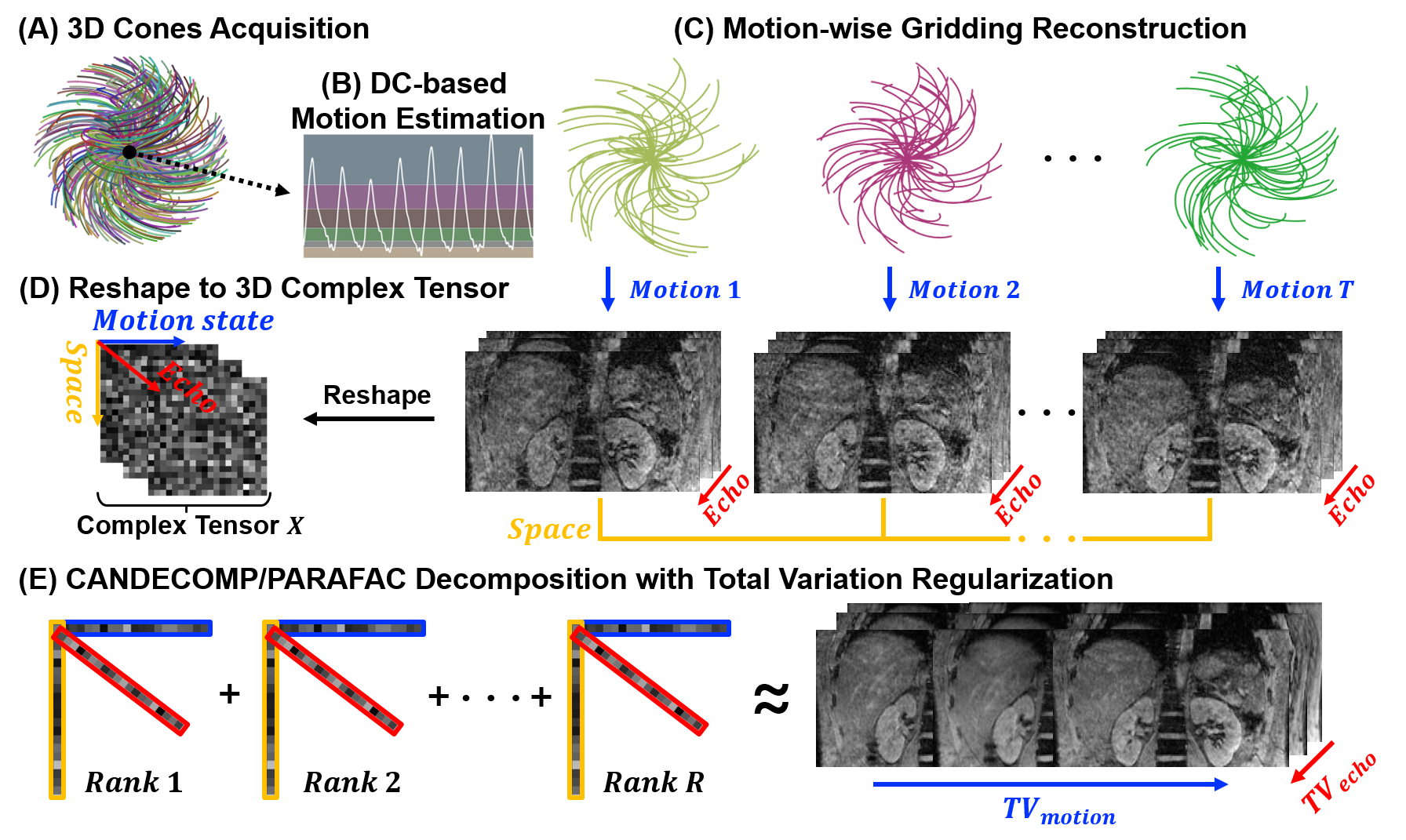}
\vspace{-4mm}
\caption{The process of constructing the input data tensor (A-D) and the proposed TV-regularized CPD (CPD-TV) method (E) for removing undersampling artifact for respiratory motion-resolved multi-echo reconstruction.} \label{fig1}
\vspace{-1mm}
\end{figure}

Let $N$ be the image size (e.g., $256 \times 256 \times 120$), $E$ be the number of TEs (e.g., TE1 to TE6), and $T$ be the number of motion states (e.g., 6 to 10). We assume that the k-space raw data is continuously acquired with non-Cartesian MRI with free breathing and motion detection is performed from the imaging data, as illustrated in Fig. \ref{fig1}. The k-space data is then sorted/binned into the motion states, and gridding reconstruction is performed on each motion state. Although the entire k-space data is acquired to meet the Nyquist sampling criterion, each motion state is highly undersampled (6-10X) due to retrospective data sorting/binning. 

Let $\mathbf{X} \in \mathbb{C}^{N \times E \times T}$ be a third order image tensor that is artifact free, and let $\mathbf{N} \in \mathbb{C}^{N \times E \times T}$ be the noise tensor that includes undersampling artifacts. Then, the image tensor $\mathbf{Y}$ from the gridding reconstruction can be written as $\mathbf{Y} = \mathbf{X} + \mathbf{N}$. The recovery of the artifact-free tensor $\mathbf{X}$ from the input tensor $\mathbf{Y}$ is the goal of this paper.

\subsection{CANDECOMP/PARAFAC Decomposition (CPD) with Total Variation (TV) Regularization}

We consider the following TV-regularized low-rank tensor decomposition/CPD for the recovery of $\mathbf{X}$.
\begin{align} \label{Eq.2}
    &\underset{\bf X}{\text{minimize}} \, \frac{1}{2}\|\mathbf{Y} - \mathbf{X}\|_F^2 + \lambda_e \cdot TV_e({\bf X}) + \lambda_t \cdot TV_t({\bf X}), \nonumber \\
    &\text{subject to } \mathbf{X} = \sum_{r=1}^R \mathbf{a}_r \circ \mathbf{b}_r \circ \mathbf{c}_r,
\end{align}
where $\| \cdot \|_F$ is the Frobenius norm, and $TV_e(\cdot)$ and $TV_t(\cdot)$ are the total variation (TV) along the echo ($e$) and motion state ($t$) dimensions defined as
\begin{align} \label{Eq.3}
    TV_e(\mathbf{X}) := \sum_{j=1}^E||\mathbf{X}(:, j+1, :) - \mathbf{X}(:, j, :)||_1
\end{align}
and
\begin{align} \label{Eq.4}
    TV_t(\mathbf{X}) &:= \sum_{k=1}^T||\mathbf{X}(:, :, k+1) - \mathbf{X}(:, :, k)||_1,
\end{align}
respectively. Note that $\mathbf{X}(:,j,:)$ and $\mathbf{X}(:,:,k)$ stand for \emph{slices} defined by fixing all but two indices, $j$ and $k$, respectively. $\lambda_e$ and $\lambda_t$ are regularization parameters. The minimization problem in \eqref{Eq.2} solves for $\mathbf{X}$ in such a way that it can be decomposed into the summation of $R$ rank-1 tensors $\mathbf{a}_r \circ \mathbf{b}_r \circ \mathbf{c}_r$, $r=1,\dots, R$, where $\mathbf{a}_r \in \mathbb{C}^N$, $\mathbf{b}_r \in \mathbb{C}^E$, and $\mathbf{c}_r \in \mathbb{C}^T$ and $\circ$ is the outer product.

\paragraph{\underline{Remark.}} The optimization problem \eqref{Eq.2} can be viewed as a special case of CPD or tensor rank decomposition where the TV regularization terms are omitted. We will henceforth refer to the proposed method (TV-regularized CPD) as CPD-TV.

\subsection{Alternating Gradient Descent}

The problem in \eqref{Eq.2} boils down to computing the factor matrices $A:= [\mathbf{a}_1, \dots, \mathbf{a}_R]$ $\in \mathbb{C}^{N \times R}$, $B:= [\mathbf{b}_1, \dots, \mathbf{b}_R] \in \mathbb{C}^{E \times R}$, and $C:= [\mathbf{c}_1, \dots, \mathbf{c}_R] \in \mathbb{C}^{T \times R}$. Unfolding/matricizing the input tensor $\mathbf{Y}$ to $Y_{(1)} \in \mathbb{C}^{N \times (TE)}$, $Y_{(2)} \in \mathbb{C}^{E \times (TN)}$, $Y_{(3)} \in \mathbb{C}^{T \times (EN)}$, we utilize the alternating gradient descent method that allows to update $A$, $B$, and $C$ one at a time. For the factor matrix $A$, we have
\begin{align}
    A \gets A - \alpha &\frac{\partial}{\partial A} (\|Y_{(1)}^\top - (C \odot B)A^\top\|_F^2 \\
    &+ \lambda_t \cdot TV_t((C \odot B)A^\top) + \lambda_e \cdot TV_e((C \odot B)A^\top)), \nonumber
\end{align}
where $\odot$ is the Khatri–Rao product and $A^\top$ is the transpose of $A$. The derivative is derived as follows:
\newcommand*{\hermconj}{^{\mathsf{H}}}
\begin{align}
    A &\gets A + \alpha \left((Y_{(1)}^\top - A(C \odot B)^\top) (\overline{C \odot B}) + \lambda_e \cdot \frac{\nabla\hermconj_e \nabla_e^{} (C \odot B)A^\top}{\| \nabla_e (C \odot B)A^\top\|_1} \right. \nonumber \\
     &\quad\quad +  \left. \lambda_t \cdot \frac{\nabla\hermconj_m \nabla_m^{} (C \odot B)A^\top}{\| \nabla_m (C \odot B)A^\top\|_1} \right),
\end{align}
where $\overline{C \odot B}$ is the complex conjugate of ${C \odot B}$, and $\nabla\hermconj$ is the Hermitian adjoint of $\nabla$. Likewise, the update equations for the factor matrices $B$ and $C$ can be similarly derived from the fact that $Y_{(2)} \approx (C \odot A)B^\top$ and $Y_{(3)} \approx (B \odot A)C^\top$.

\section{Methods}

\subsection{Human Subjects and Data Acquisition}

With IRB approval, 2 healthy volunteers and 2 patients with iron overload were recruited. These subjects were imaged with a 3T clinical MRI scanner (SignaPremierXT, GE Healthcare, Waukesha, WI). A clinically available 3D multi-echo Cartesian MRI (IDEAL-IQ) was performed with a single breath-hold as clinical reference (BH Cartesian). Subsequently, 3D multi-echo cones MRI implemented based on \cite{gurney2006design, kee2021free} was performed with free breathing (FB Cones). Imaging parameters for BH Cartesian were: Matrix size =  128(224)$\times$162(168)$\times$ 36(32), in-plane resolution = 2.9(1.78)$\times$2.3(1.78) mm$^2$, slice thickness = 6(8) mm, FA = 3$^\circ$, initial TE/$\Delta$TE/TR = 0.684(1.064)/0.6(0.86)/5.3(7) ms, rBW = 1562(1116) Hz/Px, ETL/$\#$shots = 3/2, acceleration = 4, scan time = $\sim$30 sec with a single breath-hold. Imaging parameters for FB Cones were: Matrix size = 190(226)$\times$190(226)$\times$100(120), resolution = 2$\times$2$\times$2 mm$^3$, FA = 3$^\circ$, initial TE/$\Delta$TE/TR = 0.032/1.4(1.5)/11.4(11.5) ms, $\#$TEs = 6, readout duration = $\sim$1 ms, no acceleration, scan time = $\sim$7 min.

\subsection{Construction of Input Tensor $\mathbf{Y}$}

The acquired multi-echo cones k-space data were sorted/binned into six respiratory motion states ($T=6$). To this end, a respiratory signal was estimated from the k-space center (direct current, or DC, i.e., at $k_x$=$k_y$=$k_z$=$0$) of each cone interleaf of all the coils and echoes using principal component analysis (PCA) as described in \cite{kang2022ismrm}. Respiratory motion states were then determined such that each motion state had the same number of cones interleaves for all echoes, as illustrated in Fig.~\ref{fig1}(A-C). Gridding reconstruction was performed on each set of the sorted/binned cones interleaves, and reconstructed images were reshaped to construct a 3-way input tensor $\mathbf{Y} \in \mathbb{C}^{N \times E \times T}$ as illustrated in Fig.~\ref{fig1}(D).

\begin{figure}[t!]
\centering
\includegraphics[width=\linewidth]{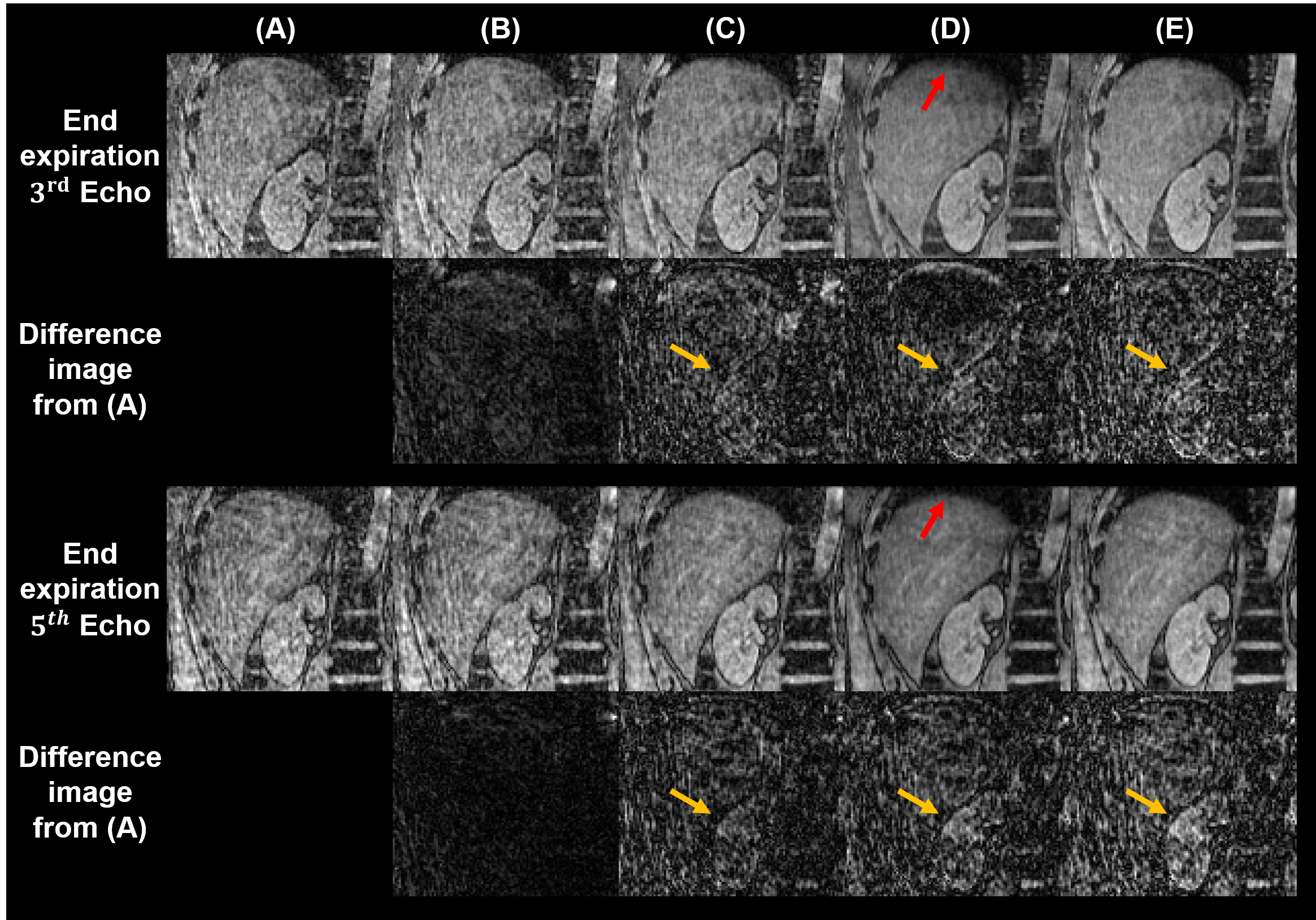}
\vspace{-4mm}
\caption{Magnitude images at TE3 and TE5 for the end-expiratory motion state of a representative healthy subject (HV \#1) and difference images between hard gating (A) and the target methods (B-E). (B) CPD with the rank of 30, (C) CPD with the rank of 13, (D) CPD with the rank of 10, and (E) CPD-TV with the rank of 13.} \label{fig2}
\vspace{-1mm}
\end{figure}

\subsection{Image Quality Assessment and R2*/QSM Measurements}

Complex-based nonlinear least squares was performed on the complex source images from CPD-TV for R2* mapping and the B0 field estimation which was then used for dipole inversion for QSM with fat-reference \cite{kang2022ismrm}. ROIs were placed on the liver tissue avoiding large vessel in coronal and axial views for R2*/QSM measurements. In addition to CPD-TV, the following image reconstruction methods were implemented and performed on 3D multi-echo cones k-space data for comparison: 1) Cartesian MRI (IDEAL-IQ) was performed with a single breath-hold as clinical reference (BH Cartesian), 2) retrospective self-gated gridding reconstruction with 17$\%$ of acceptance rate for the end-expiratory motion state (hard gating), 3) gridding reconstruction using all cones interleaves (motion averaged), 4) CS-based respiratory motion-resolved iterative reconstruction with TV \cite{feng2016xd} (motion-resolved), and 5) CPD with alternating gradient descent. Experiments were performed on an Nvidia Tesla V100 GPU, and the source code/example dataset will be available for reproducible research upon publication. 

\begin{figure}[t!]
\centering
\includegraphics[width=\linewidth]{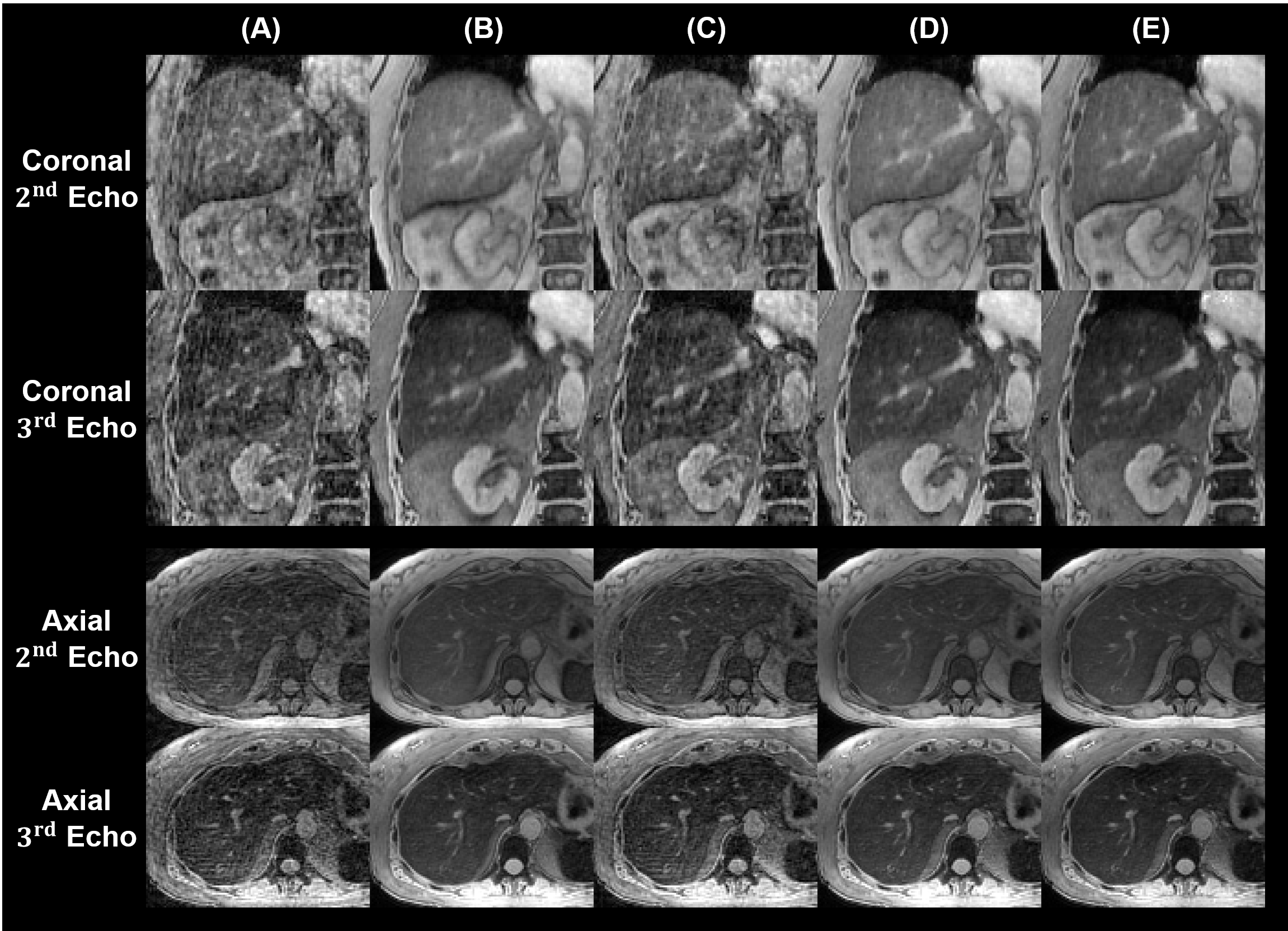}
\vspace{-4mm}
\caption{Magnitude images at TE2 and TE3 for the end-expiratory motion state of a patient (Pt \#2) with iron overload. Hard gating (A), motion-averaged (B), CPD with the rank of 13 (C), motion-resolved (D), and CPD-TV with the rank of 13 (E) reconstructions are shown in coronal and axial views.  Reconstruction times of (D) and (E) using an Nvidia Tesla V100 GPU were 4.3 hours and 1.1 hours, respectively.} \label{fig3}
\vspace{-1mm}
\end{figure}

\section{Results}
For a representative healthy volunteer, a comparison between CPD and CPD-TV for multiple choices of rank with respect to hard gating is shown in Fig.~\ref{fig2}. The undersampling artifact shown in the second and fourth rows as the difference image between hard gating (A) and CPD (B-D) was suppressed as the rank decreased from 30 to 10. The area pointed to by the red arrows in column (D) shows blurred liver/lung interface compared to (A-C), suggesting the use of a rank greater than 10. CPD-TV with the rank of 13 (E) shows excellent image quality compared to (C, D) with substantially reduced undersampling artifact shown in the difference images and the area indicated by the orange yellows – the brighter the difference image is, the more substantial the reduced undersampling artifact was.

Fig.~\ref{fig3} shows the end-expiratory motion state at TE2 and TE3 of a patient with iron overload reconstructed from CPD-TV and conventional/existing methods. The undersampling artifact in the hard gating reconstruction (A) was substantially suppressed in the motion averaged (B), motion resolved (D), and CPD-TV (E) reconstructions compared to CPD (C). The motion resolved (D) and CPD-TV (E) reconstructions presented similar image quality, with clear visualization of liver dome compared to the motion-averaged reconstruction (B). Despite similar image quality between (D) and (E), CPD-TV (E) had a reconstruction time that was approximately 4X faster than motion-resolved method (D). The computation time of CPD-TV is independent of the number of coils, whereas that of the conventional motion-resolved reconstruction gradually increases as the number of coils increases.

\begin{figure}[t!]
\centering
\includegraphics[width=\linewidth]{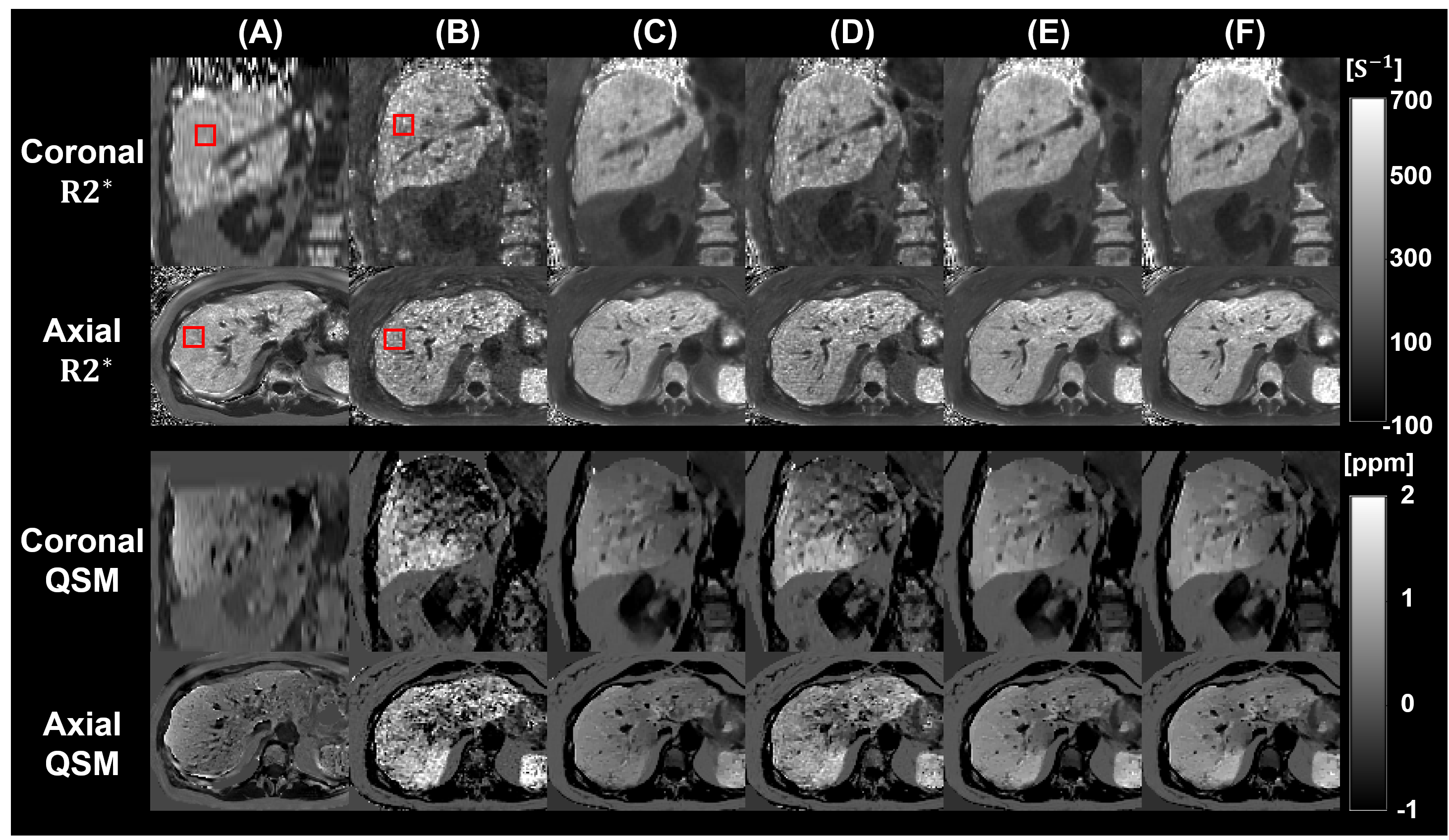}
\vspace{-4mm}
\caption{R2*/QSM reconstructions of a patient (Pt \#2) with iron overload. (A) BH Cartesian, (B) Hard gating, (C) motion-averaged, (D) CPD with the rank of 13, (E) conventional motion-resolved, (F) CPD-TV with the rank of 13. Note that the reconstructions (B-F) are from FB Cones k-space and the coronal view of (A) was interpolated along the SI direction to match the resolution of FB Cones for better visualization.}\label{fig4}
\vspace{-1mm}
\end{figure}

Fig.~\ref{fig4} shows R2* and QSM of the same patient in Fig.~\ref{fig3} with BH Cartesian as a clinical reference. Undersampling artifact in hard gating (B) and CPD with the rank of 13 (D) were manifested as noisy and over-/underestimated R2*/QSM regions in the liver compared to BH Cartesian. Motion-averaged (C) show slightly blurred image quality, and conventional motion-resolved (E) and CPD-TV (F) show similar image quality in R2*/QSM. ROIs are shown in the red squares drawn over R2* map of (A) and (B) are ROIs for R2*/QSM measurements.

\begin{figure}[t!]
\centering
\includegraphics[width=\linewidth]{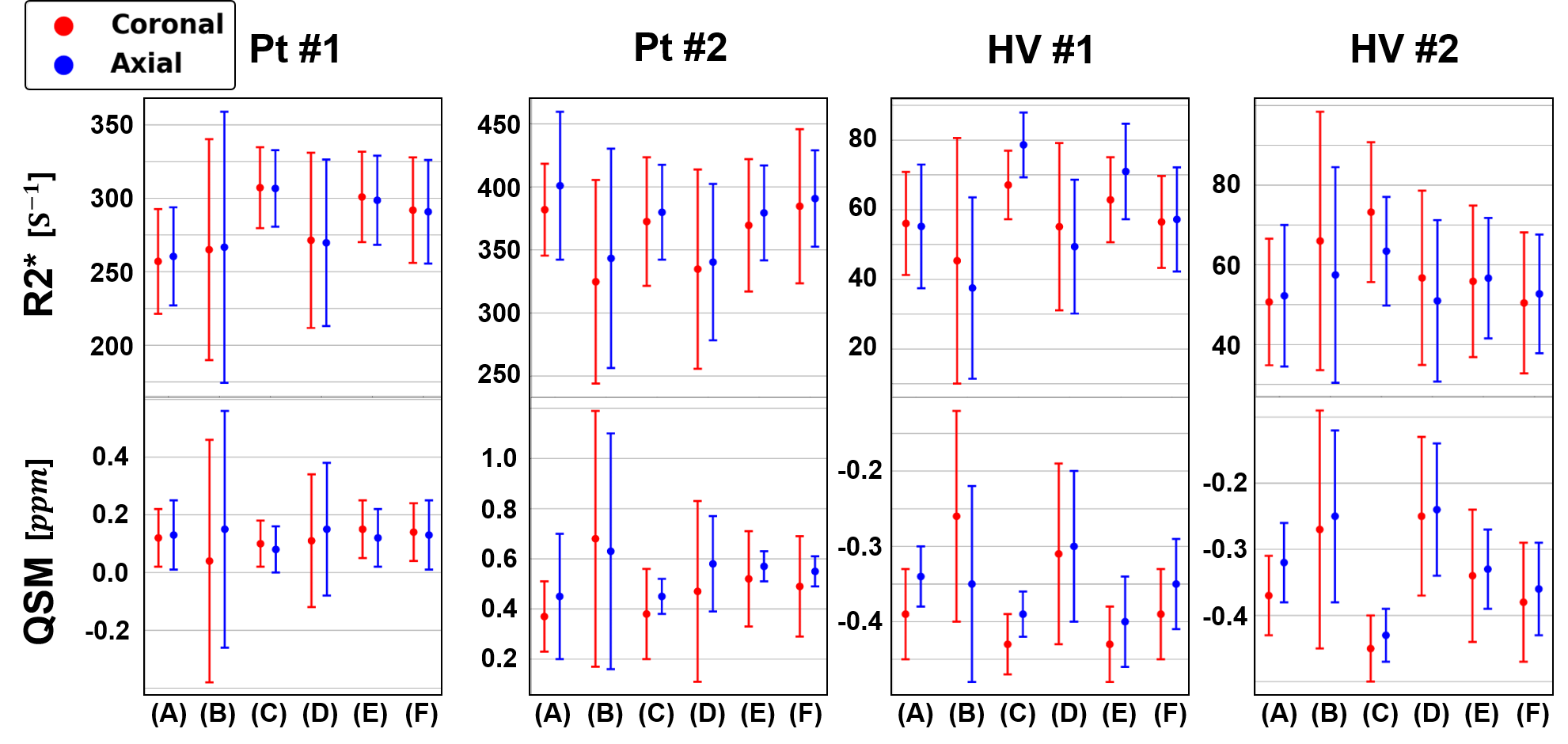}
\vspace{-4mm}
\caption{Mean and standard deviation of ROI-based R2*/QSM measurements from representative healthy volunteers (HV) and patients (Pt). (A) BH Cartesian, (B) Hard gating, (C) motion-averaged, (D) CPD with the rank of 13, (E) conventional motion-resolved, (F) CPD-TV with the rank of 13. Note that the reconstructions (B-F) are from FB Cones.}\label{fig5} 
\vspace{-1mm}
\end{figure}

For representative healthy volunteers and patients with iron overload, mean and standard deviation of ROI-based R2*/QSM measurements from BH Cartesian and FB Cones with CPD-TV and comparing methods are displayed in Fig.~\ref{fig5}. Overall, CPD-TV provided closer measurements to BH Cartesian than the comparing methods, suggesting that the potential “echo inconsistency” that arise in the conventional CS reconstruction was accounted for.

\section{Conclusion and Future Work}

We have proposed a novel tensor-based multidimensional (5D = 3D space + 1D echo evolution + 1D motion state) undersampling artifact removal method (CPD-TV). Combined with simple gridding reconstruction of highly undersampled multi-echo k-space raw data, CPD-TV has shown to successfully suppress undersampling artifacts and provide respiratory motion-resolved multi-echo images faster than conventional motion-resolved CS reconstruction, without requiring successive forward/backward 3D NUFFT operations. ROI-based R2*/QSM measurements suggest that potential “echo inconsistency” in the conventional CS reconstruction was addressed, showing similar values to BH Cartesian.

It was not until recently that the notion of low-rank tensor gained attention for multidimensional image reconstruction. We highlight some of the relevant work in terms of tensor decomposition in MRI, including dynamic cardiac T1 mapping \cite{yaman2019low} and cardiac T1, T1/T2 mapping \cite{christodoulou2018magnetic}. Future work will focus on the following areas: 1) conducting a systematic analysis of memory footprint and computational complexity, 2) exploiting data parallelism and distributed computing strategies for further speedup, 3) examining the convergence of the alternating gradient descent method for optimizing the CPD-TV objective function, and 4) applying CPD-TV to other non-Cartesian trajectories.



%
%
%

\bibliographystyle{splncs04}

\end{document}